\documentclass{iaus}
\usepackage{graphicx}
\usepackage{psfig}
\usepackage{epsf}

\title[Modelling the Chemical Evolution] 
{Modelling the Chemical Evolution}

\author[Gerhard Hensler \& Simone Recchi]   
{Gerhard Hensler \and Simone Recchi}

\affiliation{Institute of Astronomy, University of Vienna,
Tuerkenschanzstr. 17, A--1180 Vienna, Austria \\
email: {\tt gerhard.hensler@univie.ac.at, simone.recchi@univie.ac.at} }

\pubyear{2009}
\volume{265}  
\pagerange{119--126}
\jname{Chemical Abundances in the Universe: Connecting First Stars to Planets}
\editors{K. Cunha, M. Spite \& B. Barbuy, eds.}

\begin{document}
\def\cd{{\it chemo-dynamical}}
\def\SF{star formation}
\def\HI{H{\sc i} }
\def\HII{H{\sc ii} }
\def\Msun{M$_{\odot}$ }
\def\Zsun{Z$_{\odot}$ }
\def\AA{A{\rm \&}A}

\maketitle

\begin{abstract}
Advanced observational facilities allow to trace back the chemical 
evolution of the Universe, on the one hand, from local objects of 
different ages and, secondly, by direct observations of redshifted
objects. The chemical enrichment serves as one of the cornerstones
of cosmological evolution.
In order to understand this chemical evolution in morphologically 
different astrophysical objects models are constructed based
on analytical descriptions or numerical methods. 
For the comparison of their chemical issues, as there are element 
abundances, gradients, and ratios, with observations not only the
present-day values are used but also their temporal evolution
from the first era of metal enrichment.
Here we will provide some insight into basics of chemical evolution
models, highlight advancements, and discuss a few applications.

\end{abstract}

\firstsection 

\section{Introduction}

The evolution of the Universe is clearly manifested by its structure
formation and the consumption of gas to star formation, the latter
leading to the enrichment of chemical elements heavier than those
stemming from the primordial nuclesynthesis.  Knowing four facts, the
star-formation (SF) rate at any time and its integral over the Hubble
time, the initial stellar mass function (IMF), the production and
yield of each element for each particular stellar mass, in each stage
of the stellar life, respectively, would well define the interstellar
enrichment and with some SF delay also the stellar, so that an
observational trace-back of the chemical evolution (CE) of the
Universe would allow to derive one of these parameters definitely and
for different specific structures.

\begin{figure}[h]
 \includegraphics[width=3.5cm,angle=-90]{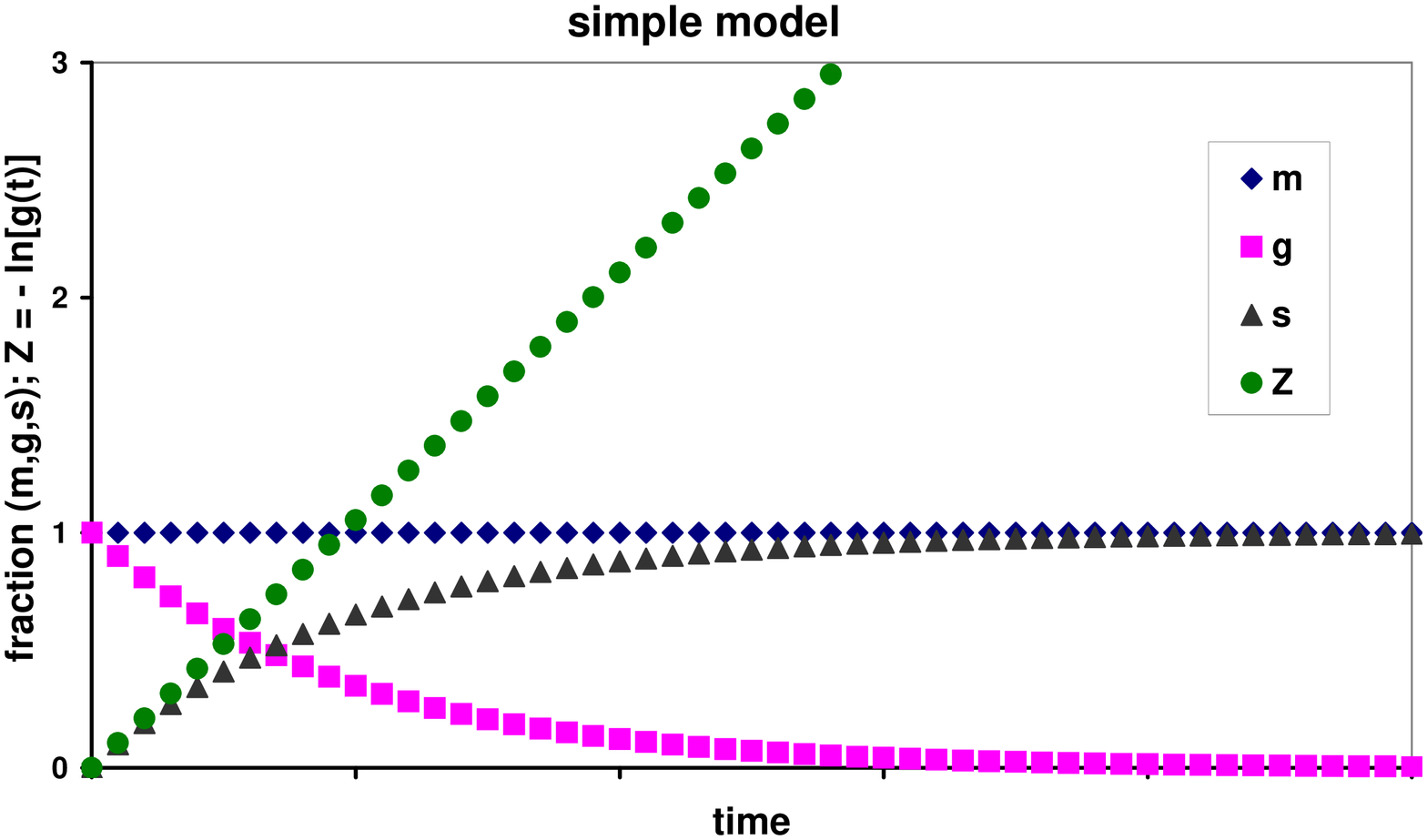} 
\hspace*{0.5 cm}
\hfill 
\includegraphics[width=3.5cm,angle=-90]{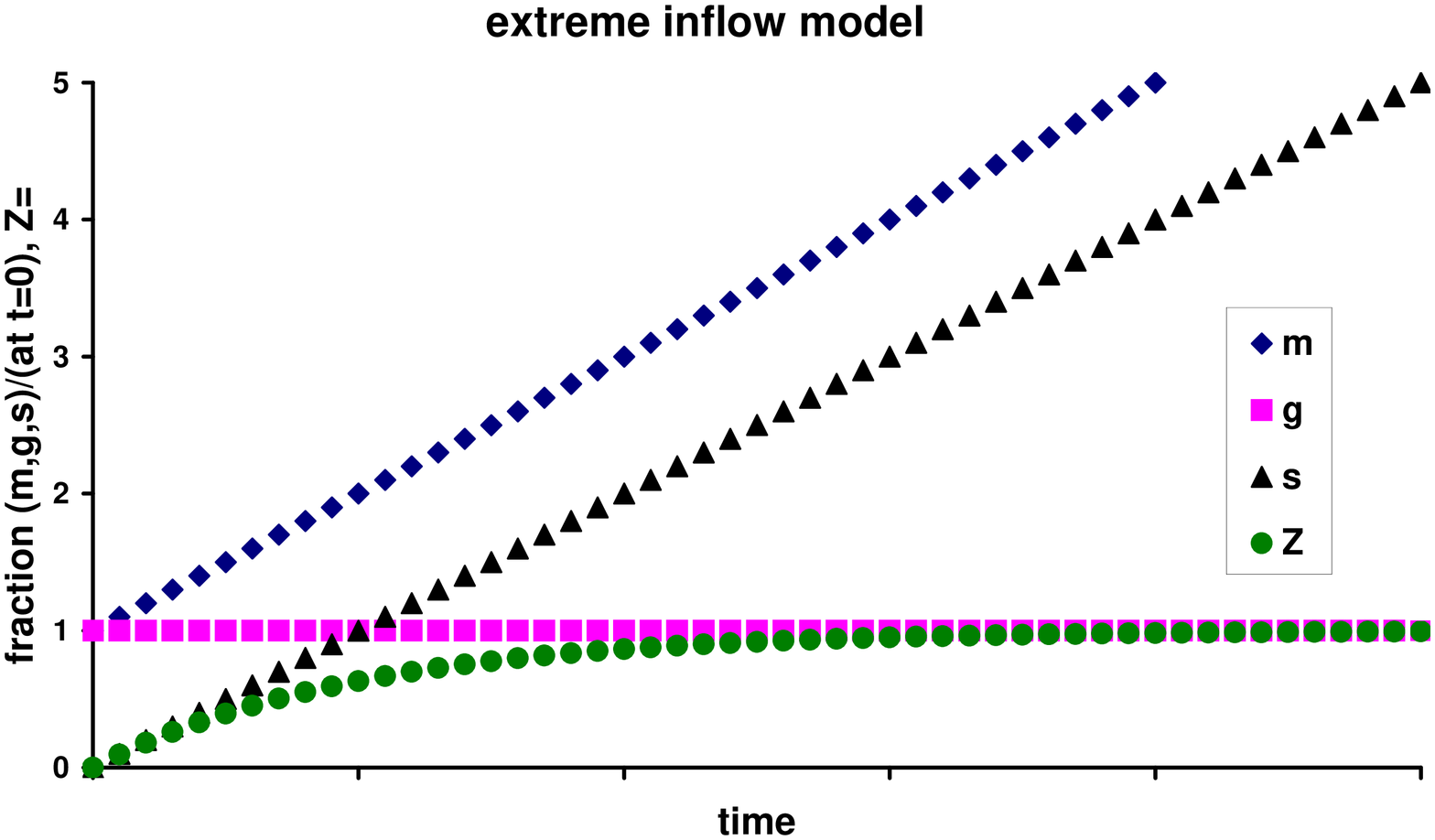} 
\caption{Evolution in arbitrary time units of 
an element $Z_i(t)$ (green dots), of total mass m (blue rhombi), 
gas mass fractions g (red squares), 
and stellar mass fraction s (black triangles) 
with respect to t=0 for a closed-box model (left) 
and for the extreme inflow of primordial gas (right) 
necessary to get the gas mass g constant.
}
   \label{fig1}
\end{figure}

Assuming that a structure like e.g. a galaxy in total or also any
galactic region under consideration behaves like an isolated volume,
called a closed box, the longer-living low-mass stars lead to increase
the stellar mass fraction continuously as well as the stellar
remnants' lock-up mass while as a consequence of SF the gas fraction
diminishes. The basic set of analytical equations for the chemical
evolution of galaxies has been formulated by numerous authors (e.g.\
\cite{tal71,pag75,tin80,pag97} and a lot more since then: see
references in the recent papers by \cite{pra08a} and \cite{rec08}). In
this simple model a temporal relation for the metallicity as $Z_i(t) =
y_i [-ln(\mu)]$ follows where $\mu = M_g(t)/M_{(g,0)}$ is the temporal
gas mass fraction and $y_i$ the yield of element i, i.e. the
metallicity release per stellar population (see left panel of
fig.\ref{fig1}).

Unfortunately, nature is not as simple and our present knowledge
has a lot of uncertainties and even gaps because of models that are
reasonably and necessarily much simpler than nature and some processes
are not yet well understood. 
A strong simplification is implied to the analytical solution by the
lifetimes of the stellar factories and thus the various delay times
of specific element releases. The discrepancy of the so-called
{\it instantaneous recycling approximation} becomes obvious
after almost 2-3 Gyrs as demonstrated in fig.\ 12 of \cite{pra08a}. 
A trace-back of element ratios through long-living stars is clearly 
representing this delay, the most prominent signature being e.g. in 
the [$\alpha$/Fe]-[Fe/H] trend of the solar vicinity.

Another simplification can be relaxed by advancing the closed box to 
open boundaries and allowing for gas infall or outflow.
Fig. \ref{fig1} (right panel) demonstrates the extreme case of gas
infall which is required to balance the gas comsumption. By this, also
the stellar Z enhancement is reduced by dilution with primordial gas.
As a further limitation the released gas is assumed to mix instantaneously 
with the local environment.

Another effect also enters into the element distribution, namely, 
the existence of different gas phases produced by means of various 
heating processes and the splitting of stellar metal injecta to these 
phases. While the SF is limited to the cool gas alone, cooling timescales 
of dilute hot gas are extremely long, by this, allowing to refrain 
characteristic elements from their immediate incorporation into newly 
formed stars. The ignorance concerning such small-scale mixing 
timescales between the different gas phases until now allows only an 
order-of-magnitude approach for chemical models. Mixing processes are
not only caused by diffusion in an otherwise static gas but triggered 
by dynamical effects. This means that last but not least basic
ingredients to understand and to model the CE is gasdynamics. 

Despite of the enormous activities all over the field of CE but because
of limited space and time in this paper, we cannot cover all aspects and 
objects under investigation of CE and also not refer to the whole bunch 
of literature, but have to pick out a few papers as representative 
and review the present state of modelling the CE of some cosmic structures 
only from the early universe to the local. We therefore apologize for 
the reasonable incompleteness.
Starting from the solar vicinity the paper focusses on some hot topics
as e.g. the galactic halo and its dwarf spheroidal (dSphs) satellite 
system finishing with dwarf gas-rich galaxies. 

Since present-day chemical models are preferentially 
performed by numerical simulations a further aspect on the quality
and reliability of models should focus on the applied numerical methods.
Since it is not the aim of this review to also address and discuss 
the validity of numerical methods of galaxy evolution, the interested 
reader is referred to a dedicated recent review by \cite{hen08}.

\section{The Milky Way components}

That our Milky Way has experienced such a CE became already
obviuos from the findings by \cite{san87} of the metallicity 
increase from halo  to thin disk stars. 
Although large-scale dynamical effects during the formation and 
evolution of galaxies, as e.g. gas accumulation by infall, galactic
winds, etc., are proven by observations, models of the galactic
CE are mostly executed for the different galactic components
in separation only and to some extent implying dynamical processes
thru analytical terms, not self-consistently but instead tuning 
them corresponding to a best fit of the observations.

\subsection{The Milky Way disk}

One of the problems within the solar neighbourhood realized at earliest 
already by Beatrice \cite{tin74} was the so-called G-dwarf problem,
the lack of low-metallicity G dwarfs. Although this problem can be
simply solved by exponentially declining infall (but also outflow or both), 
\cite{pag87} has already emphasized that additional deviations from
a constant slope in the $ln(\mu)-Z$ diagram, the yield $y$, 
point towards dynamical influences. Moreover, \cite{pra08a} 
(his fig.\ 14) shows that a metal pre-enrichment leads to a better 
fit to the stellar $Z-ln(\mu)$ distribution.

Therefore, non-dynamical CE models have been conducted with a temporal 
and radial function of a gas infall rate (see e.g. \cite{chi97b,por99}). 
Because also the SF rate is related to the gas surface density $\Sigma_g$
as $\Psi(r,t) \propto \nu\, \Sigma_g^k(r,t)$, where $\nu$ represents a SF efficiency,
the infall steers the SF. The exponent $k$ can be set to unity or adapted
to the original \cite{ken89} relation 1.4. 
Additional factors are also included, as e.g. the total mass density 
(see e.g. \cite{chi97b}, eq.4). 

The most cited model of this kind by \cite{chi97b} can fit 
the G-dwarf problem at best by two infall episodes, a short one
with 1 Gyrs timescale only and a long one of 8 Gyrs. Astrophysically 
speaking this so-called two-infall model resembles nothing exceptional
or unexpected than a short initial collapse phase and the general 
gas assembly with long-term decreasing infall rate. 
Nevertheless, also the SF efficiency as another free parameter is 
changed between both infall eposides.
Depending on the stellar sample used for the G-dwarf problem
the models fit more or less (\cite{chi97}), but a quantitative
comparison of the model abundances at present-day with solar is 
desillusionary and not well surportive for the model since it provides
not more than a qualitative tendency. The same is discernible
for the stellar age-metallicity relation (AMR) in \cite{chi97b}.

On short timescales, however, the gas accretion rate should vary locally
and affect the SF (\cite{pfl10}) and the CE (\cite{hen04a}) as well 
as the abundances (\cite{koe05}) what was also invoked from 
observations of the solar neighbourhood by \cite{kna06}.

How uncertain the deduction of relations can be and to some extent 
biased by the sample taken for studies can also be documented by a
comparison of the original AMR of stars found by \cite{twa80} 
with that derived by \cite{nor04} from a well-defined and 
complete sample of solar vicinity stars. While the traditional
results revealed the expected tendency but with very strong early
enrichment, already \cite{tad03} did not find any relation at all 
from open cluster analyses, while \cite{nor04} data yield a trend.
Showing a large hardly understandable scatter $\Delta$[Fe/H] 
by almost 0.5, even for the average AMR tendency a CE model
is hardly constructable without any metal pre-enrichment.

The evolution of radial abundance gradients is reasonable, but 
values, ranges, and the temporal behaviour are also uncertain
(see Maciel \& Costa, this volume). 
Galactic abundance gradients that are measured from \HII regions as 
usually by the standard method of collisionally excited 
emission line (CEL, \cite{sha83}) or recently used recombination lines (RL)
in the infrared (\cite{est05}) can only yield the present-day abundances
in the ISM.  Over years of determinations the slope has not only 
been reduced, e.g. for log(O/H) from -0.8 per kpc (\cite{sha83}) 
to -0.44 (\cite{est05}), but also a flattening of the gradients
in the innermost radii and also outside the solar circle (\cite{vil96})
became available from RLs. From massive stars one can derive the same:
the chemical abundances of the ISM from which they are born most recently.

While \cite{mac03} could derive a temporal flattening of the oxygen 
gradient over the past 9 Gyrs from -0.11 to -0.06 dex/kpc unsing Planetary 
Nebulae, from stars of different ages \cite{nor04} could even attribute
a positive slope to disk stars older than 10 Gyrs, and again 
a slight flattening for younger stars from almost -0.1 dex/kpc to 
-0.077 since 1.5 Gyrs. 

While the present-day abundance gradients from \HII regions hold 
for all the determined elements, i.e. C, N, O, and $\alpha$ 
elements, their different radial slopes need explanation. 
Interestingly, the abundance ratio C/O shows also a radial decline 
what means that C is more abundant in O-rich regions. 
As explanation \cite{cari05} proposed that this behaviour is caused 
by a metal-dependent C yield. On the other hand, also the O yield is metal
dependent in the sense that at lower metallicities a weaker wind by massive
stars allow a longer shell-burning production of O (\cite{mae92}). 
\cite{ces09} modelled the chemical evolution of bulge and disk stars with 
metal-dependent yields by \cite{mey02b} and could by this also reproduce 
the enhancement of C/O in metal-rich regions like the bulge.

\begin{figure}[ht]
 \includegraphics[width=13cm]{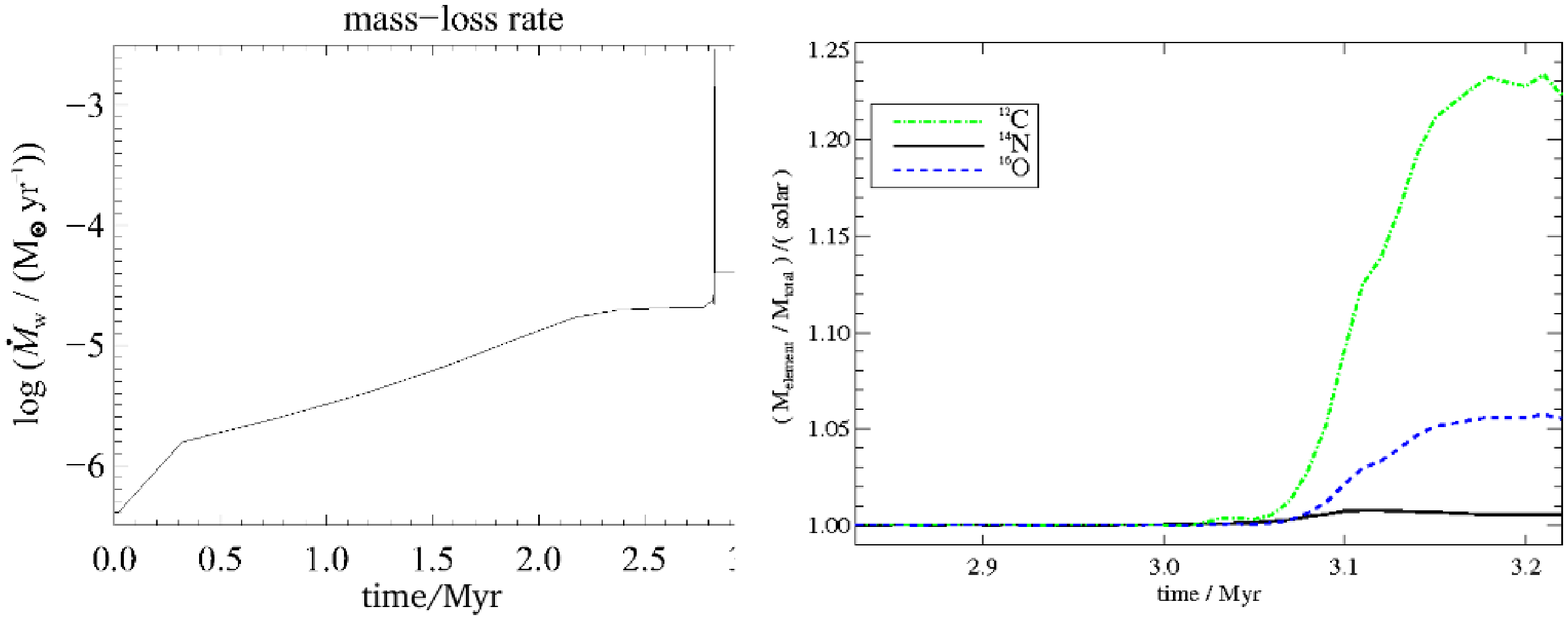} 
\hfill
\caption{Evolution of a 85 \Msun star until its explosion as supernova typeII. 
{\bf Left panel:} Wind mass loss adapted to evolutionary models by \cite{sch92}. 
{\bf Right panel:} Time-dependent abundances of $^{12}$C, $^{14}$N, 
and $^{16}$O in the warm \HII region gas after the onset of the WN phase 
at 2.83 Myrs. See \cite{kro06b}.
}
   \label{fig2}
\end{figure}

For the determination of C and O in \HII regions, another effect has 
to be studied seriously: self-enrichment by Wolf-Rayet stars
(\cite{kro06}). 
At the end of its lifetime at $t = 3.22$ Myr a 85 \Msun star has 
supplied 0.28 \Msun of $^{14}$N, 13.76 \Msun of $^{12}$C, and
11.12 \Msun of $^{16}$O, which are contained in the combined stellar
wind bubble/\HII region. Since N was released at first in the WN stage 
it increases slightly in this period and is thereafter diluted by the 
N-poor gas feed. 
These facts are discernible in the hot gas by a first rise and a subsequent 
decrease of the N abundance (\cite{kro06}) after the transition to the 
WC stage when C and O are released. The C content in the hot gas
increases steeply and reaches an overabundance of 38 times solar
while the enrichment with O is weaker. 
Due to turbulent mixing of the hot gas with the photo-evaporated shell, 
the warm \HII gas is also enriched but at most for C by not more than a
factor of 1.22 (Fig. \ref{fig2}, right panel). Since the stellar mass
range of the Wolf-Rayet phase increases with metallicity, the negative slope
of the C/O ratio within the galactic disk can be attributed to this effect.

In a recent study \cite{col09} found that the chemical properties of the galactic
disk are at best reproduced by models with an infall law derived from
cosmological but purely Dark Matter simulations. In reality, the disk settling 
timescale is not a simple stellar dynamical one but inherently determined by 
its thermal and turbulent energy budget causing a delayed thin-disk
formation following the thick disk (\cite{bur92}).

\subsection{Multi-zone models}

As already implied as advancement of simple CE models one can even account 
for gas exchanges through the boundaries of limited regions under 
consideration by simply adding a corresponding time-dependent 
(source or sink) term to the gas-mass and metallity equations.  
This was e.g. done in several above-mentioned models with gas infall 
to the galactic disk. Nonetheless, no simple model exists that also
accounts for outflow from the disk. 
A first step was the exploration of the effect of the galactic fountain 
model on radial metallicity gradients by \cite{spi09}. The idea is, that 
the disk gas pushed-up by supernovae (SNe) and redistributed while falling back on ballistic paths has to mix with the disk gas within the site of impinge 
and affects the abundance gradient of the galactic disk. 
As general results they found that neither the ballistic delay nor the
cooling delay of this fountain-driven mixing process alter
the metallicity and the radial oxygen gradient.

As well as in this approach also e.g. for infall models, authors preferably
avoid the complexity of a proper hydrodynamical treatment combined with the 
chemical and thermal descriptions by means of an analytical ansatz. 
From gas infall and expected radial gas flows, the above mentioned models 
mimic the first process by radius-dependent functions, but drop the latter 
effect. 
In order to differentiate between radial disk regions and to follow
radial flows, multi-zone models were  developed e.g. that devide the 
galactic disk into rings and allow for a radial gas flux, are artificially
parametrized due to positive flows characterizing the inside-out evolution 
of the disk (\cite{por00}) or are consistently determined by the gas 
pressure gradient. 

Different approaches were performed by coupling structural parts of the 
Milky Way, e.g. halo and disk, and dealing with mass exchange 
between the regions (\cite{fer92,fer94}). By fine-tuning not only the
coefficients for mass exchange arbitrarily, but also further parameters,
as e.g. SF efficiency, etc., rough quantitative agreements with gas
and stellar mass, abundances, and their radial gradients in the disk could 
be achieved.

\subsection{Chemo-dynamics}

For all these simple-type CE models it is inherent that their results
in comparison with observational facts can provide a first insight
into amount and timescales on which non-local effects have played a
role during the evolution. Nonetheless, they are all devoid of
self-consistency according to the acting processes and the resulting
dynamics. Two major sources exist, driving the gas dynamics, namely,
gravitation and stellar energy release. Plasmaphysical processes
according to heating and cooling lead to different gas phases that
have not only different energetics but also behave dynamically
differently and interact by multiple processes as e.g. by drag forces,
shock waves, and turbulence as well as energeticly e.g. by heat
conduction. Moreover, this means that also SF and thermal state of the
ISM have to be treated consistently.

A first-order approach into this direction is provided by multi-phase CE
models (also implied in \cite{fer92,fer94}) where, however, also the
interaction terms in a set of equations are not self-consistently determined
but arbitrarily applied.  

How crucial an appropriate representation of the ISM and its processes is,
will be demonstrated in a few of the next sections. Two main ingredients
can already be established here: 
1) Cool/warm and hot gas phases are coupled energeticly by heat conduction,
leading to a self-regulation of SF (\cite{koe98}), local mixture of hot 
and cool gas
and, by this, local metal enrichment of cool gas with SN elements.
2) The dynamical coupling of both gas phases by drag and mass loading due to
evaporation hampers the outflow and enhances the cooling of the hot gas.

\begin{figure}[ht]
\vspace{1.2cm}
\begin{center}
\includegraphics[width=130pt, angle=-90]{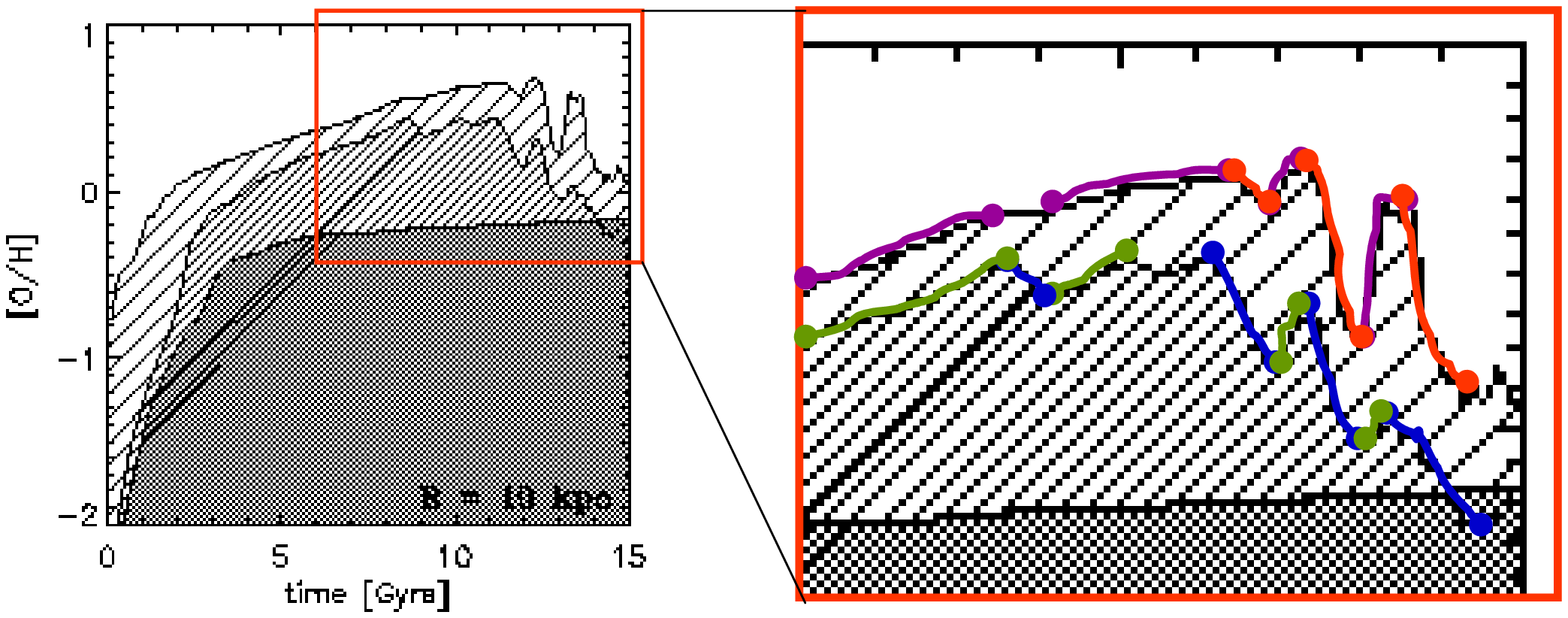} 
\end{center}
\vspace*{0.2cm}
\caption{Metallicity evolution of three different components, 
cool/warm gas (CM, dark hatched), 
hot intercloud medium (ICM, light hatched), 
and stars (grey) on average of a Milky Way-type \cd\ model 
(\cite{sam97}) at 10 kpc galactocentric distance within the disk. 
The right panel is a cut-out of the left one demonstrating how the 
metallicity Z is affected by different processes: 
1) Due to infall of primordial gas, Z$_{CM}$ decreases. 
2) Z$_{ICM}$ increases due to supernova-expelled metals and is
always higher than Z$_{CM}$. 
3) The Z$_{ICM}$ decreases therefore, when CM evaporates by heat conduction. 
4) Condensation of hot gas onto cool clouds increases their Z. 
The stellar Z represents an average over stars of all ages and is 
therefore smaller than Z$_{CM}$. (from \cite{sam94}).
}
   \label{fig3}
\end{figure}

Up to now as the optimal grid code one can consider the further development
of the \cd\ scheme by \cite{sam97b} to 3D and with the stellar dynamics
for the stars. In addition, a cosmologically growing DM halo is included
into the simulations by \cite{sam03}. These models contain all the crucial
processes of SF self-regulation by stellar feedback, multi-phase ISM,
and temporally resolved stellar components according to the \cd\ 
prescription (\cite{hen03}). They cannot only trace 
the formation and evolution of the disk galaxies' components but also of
characteristic chemical abundances and are until now the best
self-consistent evolutionary models of disk galaxies because the disk
formation is included into the global temporal galaxy evolution.

\subsection{The Milky Way halo}

Until half a decade ago only poor knowledge on the element abundances
of the Milky Way halo stars was available. With the wealth of new
class 8-10m telescopes, spectroscopy of faint and distant halo stars
has changed our picture fundamentally. While in the era of
\cite{san87} a halo metallicity [Fe/H] not below -4 could be found and
only a few below -3, raising doubts on the first population of stars
in the halo and their chemical fingerprints, recent detections of
hyper metal-poor stars (\cite{bee05}) in the galactic halo and their
peculiar element abundances (see e.g. \cite{fre05}) opened a wide
field of new activities, namely, modelling towards understanding the
zero metallicity nucleosynthesis and studying the formation of the
halo. This is necessary in order to decide between the two preferred
formation scenarios of the Milky Way: the monolithic collapse model
(\cite{egg62}) and the accretion model (\cite{sea78}).  A general word
of clarification of the two terms should be done here: Even if the
accretion model has formed an inhomogeneous halo (\cite{arg00,oey00})
its mixing of infalling gas with stellar ejecta lead to a settling of
more or less homogenised gas with low angular momentum into the bulge
and with larger one into a disk. Due to the higher gas column density
in the innermost disk due to its smaller area the mostly invoked
inside-out evolution in the disk is quite plausible.

Although a dichotomy of halo stars, the inner 
halo vs. the outer halo (\cite{caro05}), makes a dissipative formation
of the inner halo probable but invokes a dissipationless accretion 
process for the outer halo stars, it is still under debate whether part
of the present-day dwarf spheroidal galaxies (dSphs) could have served 
as its building blocks. Although most of the dSphs have formed their
stellar population in an early short, but active SF epoch, the
chemical signatures of most of the dSph stars still alive are distinct 
from the stars of each kinematic component of the Milky Way (\cite{ven04}). 

Even the most recent investigation by \cite{pra08b} which is sold 
as a clear proof that dSphs have served at least partly as building blocks 
of the Milky Way halo, lacks of consistency: SF history, baryonic mass content 
and metallicity have been assumed from the 4 most prominent present-day dSphs, 
and the accretion timescale was set to half a billion years only.

Since even the CE of the dSphs is not yet conclusively understood, 
in general, models have to be advanced for this morphological galaxy type.

\section{Dwarf galaxies}

\subsection{The Milky Way dwarf spheroidal satellite system}

Since the origin and the Dark Matter content of the galactic system of
dSph satellites is basically questioned by some authors
(\cite{met09}), general problems are: Can low-mass systems survive an
intense SF epoch (as expected at the formation)?

From compilations by \cite{dek03} and \cite{gre03b} dSphs follow a
mass-metallicity relation. As an extreme case \cite{hen04bb} have
performed spherical low-mass galaxies by means of \cd\ simulations in
order to study the galaxy survival, SF rates, gas loss, and (final)
metallicity.  They could demonstrate that due to the SF
self-regulation only short but vehement initial SF epochs occur and
lead to mass-dependent gas loss.  Nonetheless, the dwarf galaxies
(DGs) remain gravitationally bound with the further issue that more
cool gas survives than it is observed, but it forms a halo around the
visual body. Although the stellar energetic feedback is the driving
mechanism to expel the gas, its effect is not as dramatic as obtained
in semi-analytic models (\cite{sal08}) and the amount of unbound mass
is considerably lower. To get lost, this gas has to be stripped off
additionally (\cite{gre03}) what probably happens because of the ram
pressure of the galactic halo gas or of tidal stripping
(\cite{rea06}). Otherwise it can bounce back to the DG and produce
subsequent events, from a second SF epoch even up to SF
oscillations. The external gas reservoir around the Scl dSph
(\cite{bou03}) might whitness this effect.  While the models make some
intermediate-age stars, most stars are made in less than 1 Gyr after
formation. The dependence of stellar metallicity on the galaxy mass is
a quite well confirmed observational relation.

Because of its simplifying one-dimensionality and the neglect of
environmental effects like e.g. tidal effects, external gas pressure,
gas inflow, etc. more complex models are required. The fascinating
wealth of data and their precision on stellar ages and kinematics, on
their chemical abundances and tidal tails of dSphs (for most recent
reviews see e.g. \cite{koc09} and \cite{tol09}) have triggered
numerous numerical models. Although all of them up to date are
advanced by 3D hydrodynamics (see e.g. \cite{mar06} and \cite{rev09}),
they still lack of the same aforementioned environmental agents. The
most comprehensive paper by \cite{rev09} e.g. simulates a bunch of DG
models with the method of smooth-particle hydronamics (SPH), but again
all of them in isolation. In their models mainly also sufficient gas
mass is retained and can fuel further SF epochs if it would not be
stripped of by ram pressure or tidal forces, as the authors mention.
Those models that fit the presently best studied dSphs Fnx, Car, Scl,
and Sex at best, are than assumed as test cases for further
exploration.  Although their results do not deviate too much from the
further observational data, in addition to the already mentioned
neglects, three further caveats must be mentioned: 1) If models are
selected according to agreement with one or two observed structural
parameters, it is not surprising if also other values would not
deviate significantly.  2) The numerical mass resolution of the SPH
particles is too low to allow quantitative issues of galactic winds,
heating and cooling, etc.  3) Because of the single gas-phase
description released metals are too rapidly mixed with the cool gas
and the metal-enrichment happens too efficiently. Despite these facts,
with appropriate initial conditions always models in agreement to
observations can be found.

Although the advancement to a two-phase ISM by an SPH code is not
trivial and inserts various numerical problems, but is not impossible
(\cite{ber03,har06,sca06}), such treatment would be absolutely
necessary in order to achieve reliable results.  In addition the \cd\
interaction proceeses must be implied.

\subsection{Dwarf Irregular Galaxies}

Dwarf irregular galaxies (dIrrs) are characterized by large gas
fractions, often an ongoing SF and low metallicities. Until $\sim$ 10
years ago it was believed that at least some of them were forming
stars now for the very first time. Nowadays it is evident that even
the most metal-poor ones (like IZw18) contain stars at least 1 Gyr old
(\cite{momany05}). This means that SF should have proceeded for at
least a few Gyr in dIrrs, albeit at a low intensity. The low intensity
of the SF in DGs is the best way to explain their chemical
characteristics, like for instance the low [$\alpha$/Fe] ratio. As
explained in the introduction, the [$\alpha$/Fe] vs. [Fe/H] plot is
representative of the different delay in production of
$\alpha$-elements (mostly produced by the short-living massive stars)
and the iron (2/3 of which comes from longer-living binary systems
originating Type Ia SNe). If the SF duration in a galaxy is very
short, Type Ia SNe do not have enough time to restore Fe into the ISM
and most of the stars will be over-abundant in $\alpha$-elements
compared to Fe. The low average [$\alpha$/Fe] ratios in dIrrs
(compared to large galaxies) is a hint of a long-lasting (presumably
very mild) SF in these galaxies (\cite{ia02}; \cite{lm04}). The same
trend of [$\alpha$/Fe] (decreasing as we move towards dwarf galaxies)
can be obtained by varying the IMF (\cite{recc09}), but the hypothesis
of a low SF efficiency in dwarf galaxies is important also to explain
the downsizing (\cite{cowie}), namely the trend of having older
stellar populations as we move towards more massive galaxies,
indicative of a shorter (but more intense) duration of the SF in large
galaxies.

On the other hand, most people believe that a fundamental role in the
chemical evolution of dIrrs is played by galactic winds. In fact,
these systems are characterized by shallow potential wells and less
energy is required to extract gas from them. Galactic winds have the
effect of reducing the metallicity of a galaxy in comparison with the
one predicted by the closed box model. However, detailed numerical
simulations (\cite{db99}; \cite{recc06}) have shown that galactic
winds are not very effective in removing gas from a galaxy. In fact,
galactic winds develop vertically (along the direction of steepest
pressure gradient) whereas the horizontal transport along the disk
(where most of the gas lie) is very limited. On the other hand, the
metals freshly produced during a starburst can be easily carried out
of the galaxies through a galactic wind (which will be therefore
metal-enhanced) and this, too, contributes to keep the metallicity of
the galaxy low. Although this overall picture is almost widely
accepted, the effect of galactic winds depends very sensibly on
parameters like galaxy structure and ISM properties. In particular,
the presence of clouds can hamper the development of galactic winds
while in the meantime, through evaporation of metal-poor gas, reducing
the average metallicity of the galaxy. Detailed numerical simulations
(\cite{rh07}) show that the leakage of metals is not prevented by the
presence of clouds and that the final metallicity is a few tenths of
dex lower than in models without clouds.




\section{Concluding Remarks}

Because of its importance for understanding the evolution of baryonic 
structures in the Universe, their chemical evolution is one of the main
focusses of galaxy investigations. For simplicity, numerous
analytical approaches have been performed for different morphological
units. They can provide a first insight into the temporal
and spatial element enrichment by stellar nucleosythesis products.

Since dynamical effects of different gas phases affect the element 
enrichment differentially leading to a redistributions of those elements,
but because also plasmaphysical processes as e.g. heat conduction and
turbulence allow for a mixing of gas, more complex dynamically and chemically
coupled descriptions have to be developed for advanced models.
The development and application of multi-phase so-called \cd\ descriptions
are on the way.

How important the implementation of the multi-phase character of the ISM is,
can be simply understood by the following basic considerations of its 
inherent nature: In a low-mass galaxy the stellar energy release can
overcome the binding energy of the whole gas reservoir and the gas
is totally lost - and by this also the freshly produced metals. 
Semi-analytical (\cite{sal08}) or single-phase hydrodynamical (\cite{mcl99})
models largely overestimate the gas loss due to the instantaneous energy 
mixture, but to some extent also because of an arbitrarily high SF rate. 
Although the first-order 1D \cd\ approach (\cite{hen04b}) even
overestimates the momentum transfer by the drag exerted from hot gas flows
to the cool clouds and leads to an overly more efficient gas removal 
than in 2D or 3D \cd\ models e.g. of dIrrs (\cite{hen99}), the model
description is still closer to reality because it allows the 
outflow of hot gas, while hampered by mass loading and drag of cool 
infalling intergalactic clouds. The amount of expelled 
metals is reduced (\cite{rh07}). 

On massive galaxy scales, as e.g. the evolution of giant ellipticals
(gE) (\cite{pip06}), however, this picture can turn. When hot SNeII
gas of $10^6-10^7$ K mixes with a sufficiently high amount of 10$^4$ K
and cooler gas, a rough estimate gives that with a SF efficiency of
10\% and a Salpeter IMF about 10\% of the stellar mass are returned by
massive star ejecta, i.e. winds + SNeII, what is 1\% of the initial SF
gas. Accordingly, the temperature of the mixed gas then amounts to
$2.1\cdot 10^4$ - $1.2\cdot 10^5$ K, i.e. always in the range of
shortest cooling time, and, by this, remains always bound. A
collapsing massive galaxy is not stopped and metal-enriched gas is
completely following the collapse.  Lateron, during the evolution this
changes due to the gas consumption so that the SN gas dominates and
gas at above $10^6$ K has reasonably to re-expand (\cite{pip06}). It
is not surprising that out of a model grid any model will be found to
fit the observations at best, but nonetheless the validity of its
issued parameters must be questioned. Again the simple 1D \cd\ model
(\cite{the92}) demonstrates that a metallicity of 1-2 times solar can
be reached within 1 R$_e$ of a gE, while the hot gas with 3-4 times
solar metallicity extends out to 10 R$_e$.

The here discussed considerations aim at sensibilizing the reader to
the validity of CE models. Finally, we wish to mention that the
considerations of CE of tidal-tail DGs, of the intra-cluster medium,
of Ly-alpha structures in the early Universe, and furthermore, are
also in the focus of crucially important studies and provide further
details and to some extent initial and boundary conditions for the CE
of galaxies.

Moreover, one should also keep in mind that the stellar yields applied 
to CE models are still uncertain and change due to advanced stellar
evolutionary models, taking into account stellar rotation 
(see e.g. \cite{chi03,hir05,mey06}), binarity (\cite{ded04}), stellar winds,
etc. 
\\

{\bf Acknowledgement:} The authors wish to thank A. Hren for providing
some figures, the symposium organizers for their invitation to this
review, and the University of Vienna for travel support.

\end{document}